\documentclass[sigconf]{acmart}

\usepackage{color}
\usepackage{float}
\usepackage{subcaption}
\usepackage{xspace} 
\usepackage{graphicx}
\usepackage{multirow}
\usepackage{array}
\usepackage{enumitem}
\usepackage{ctable} 
\usepackage{longtable}

\AtBeginDocument{%
  \providecommand\BibTeX{{%
    \normalfont B\kern-0.5em{\scshape i\kern-0.25em b}\kern-0.8em\TeX}}}

\copyrightyear{2022} 
\acmYear{2022} 
\setcopyright{rightsretained} 
\acmConference[UIST '22]{The 35th Annual ACM Symposium on User
Interface Software and Technology}{October 29-November 2, 2022}{Bend,
OR, USA}
\acmBooktitle{The 35th Annual ACM Symposium on User Interface
Software and Technology (UIST '22), October 29-November 2, 2022,
Bend, OR, USA}
\acmDOI{10.1145/3526113.3545661}
\acmISBN{978-1-4503-9320-1/22/10}

\newcommand{\bolderandunderline}[1]{\textbf{\underline{#1}}}

\newcommand{\systemname}{Wigglite\xspace}
\newcommand{\systemnameInFull}{\bolderandunderline{W}iggling for
\bolderandunderline{I}nformation
\bolderandunderline{G}athering and
\bolderandunderline{G}enerating
\bolderandunderline{L}ightweight
\bolderandunderline{I}mpressions for
\bolderandunderline{T}riage and
\bolderandunderline{E}ncoding\xspace}
\newcommand{\basesystemname}{\textsc{Skeema}\xspace}

\newcommand{\stackoverflow}{Stack Overflow\xspace}

\newcommand{\javascript}{JavaScript\xspace}
\newcommand{\typescript}{TypeScript\xspace}
\newcommand{\css}{CSS\xspace}
\newcommand{\html}{HTML\xspace}

\newcommand{\code}[1]{\texttt{#1}}
\newcommand{\userquote}[1]{``\textit{#1}''}

\begin{document}

\title{
    \systemname: Low-cost Information Collection and Triage
}


\author{Michael Xieyang Liu}
\affiliation{%
  \country{Carnegie Mellon University}
  }
\email{xieyangl@cs.cmu.edu}

\author{Andrew Kuznetsov}
\affiliation{%
\country{Carnegie Mellon University}
  }
\email{adkuznet@cs.cmu.edu}

\author{Yongsung Kim}
\affiliation{%
   \country{Carnegie Mellon University}
  }
\email{yongsung@cmu.edu}

\author{Joseph Chee Chang}
\affiliation{%
   \country{Allen Institute for AI}
  }
\email{josephc@allenai.org}

\author{Aniket Kittur}
\affiliation{%
   \country{Carnegie Mellon University}
  }
\email{nkittur@cs.cmu.edu}

\author{Brad A. Myers}
\affiliation{%
   \country{Carnegie Mellon University}
  }
\email{bam@cs.cmu.edu}

%
%
%
%
%

\renewcommand{\shortauthors}{Liu, Kuznetsov, Kim, Chang, Kittur, and Myers}

\begin{abstract}
  Consumers conducting comparison shopping, researchers making sense of competitive space, and developers looking for code snippets online all face the challenge of capturing the information they find for later use without interrupting their current flow. In addition, during many learning and exploration tasks, people need to externalize their mental context, such as estimating how urgent a topic is to follow up on, or rating a piece of evidence as a ``pro'' or ``con,'' which helps scaffold subsequent deeper exploration. However, current approaches incur a high cost, often requiring users to select, copy, context switch, paste, and annotate information in a separate document without offering specific affordances that capture their mental context. In this work, we explore a new interaction technique called ``wiggling,'' which can be used to fluidly collect, organize, and rate information during early sensemaking stages with a single gesture. Wiggling involves rapid back-and-forth movements of a pointer or up-and-down scrolling on a smartphone, which can indicate the information to be collected and its valence, using a single, light-weight gesture that does not interfere with other interactions that are already available. Through implementation and user evaluation, we found that wiggling helped participants accurately collect information and encode their mental context with a 58\% reduction in operational cost while being 24\% faster compared to a common baseline.
\end{abstract}

\ccsdesc[500]{Human-centered computing~Interactive systems and tools}

\keywords{Sensemaking, Mental models, Interaction technique}


\begin{teaserfigure}
  \vspace{-1mm}
  \includegraphics[width=1\linewidth]{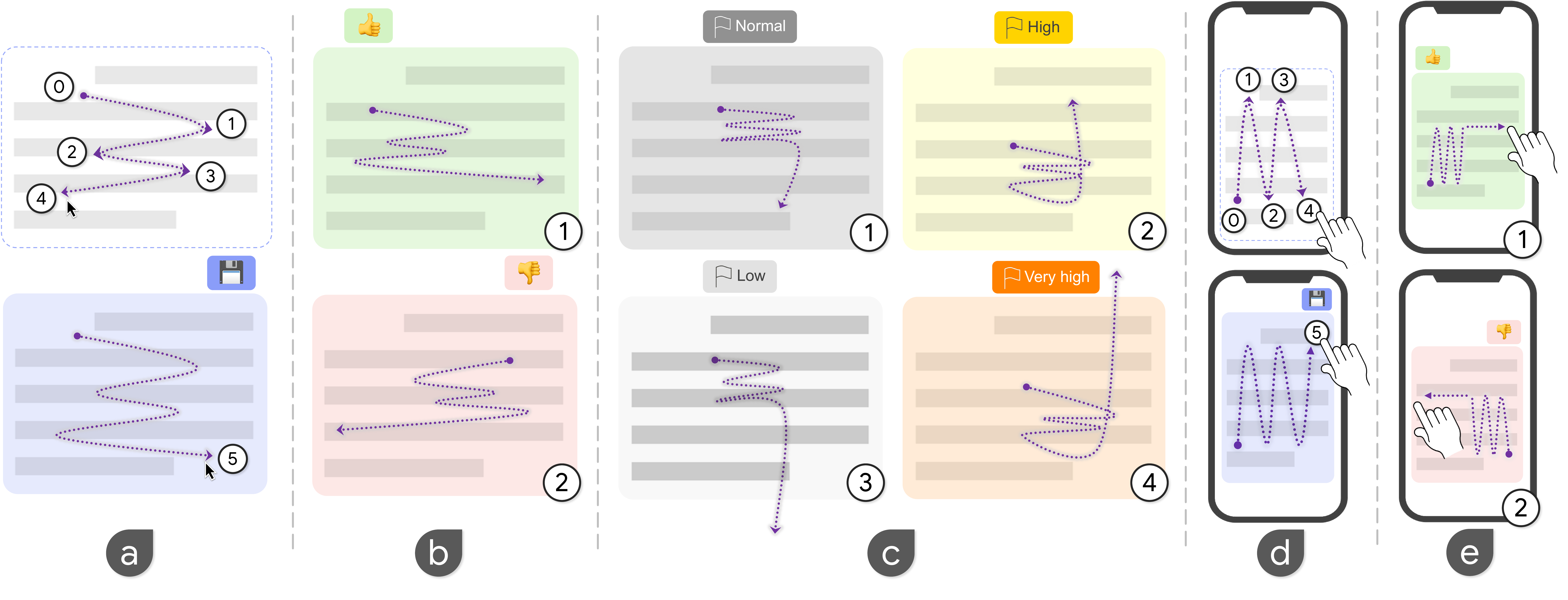}
  \vspace{-9mm}
  \caption{
  We introduce the ``wiggling'' technique: rapid back-and-forth movements of a mouse pointer on desktop (a) or a finger on mobile devices (d) that do not require any clicking to perform, yet are sufficiently accurate to select the desired content, while at the same time supporting an optional and natural encoding of valence rating (positive to negative) (on desktop: b1-2, on mobile: e1-2) or classification of priority (to facilitate triage) (c1-4) by ending the wiggle with a swipe in different directions.}
  \label{fig:wiggle-detailed-graph}
\end{teaserfigure}

\maketitle

\section{Introduction}
From consumers researching products, to patients making sense of medical diagnoses, to developers looking for solutions to programming problems, people spend a significant amount of time on the internet discovering and researching different options, prioritizing which to explore next, and learning about the different trade-offs that make them more or less suitable for their personal goals \cite{chang_mesh_2020,liu_unakite:_2019,liu_reuse_2021,kittur_costs_2013,chang_searchlens_2019,chang_tabs_2021}. For example, a YouTuber seeking to upgrade her vlogging setup may learn about many different camera options from various online sites. As she discovers them, she implicitly prioritizes which are the most likely candidates she wants to investigate first,  looking for video samples and technical reviews online that speak either positively or negatively about those cameras. Similarly, a patient might keep track of different treatment options and reports on positive or negative outcomes; or a developer might go through multiple \stackoverflow and blog posts to collect possible solutions and code snippets relevant to their programming problem, noting trade-offs about each along the way.

While the number of options, their likely importance, and evidence about their suitability can quickly exceed the limits of working memory, the high friction of externalizing this mental context means that people often still keep all this information in their heads \cite{hinckley_informal_2012,marshall_saving_2005,tashman_active_2011,liu_unakite:_2019,chang_supporting_2016}. Despite the multiple tools and methods that people use to capture information, such as copying and pasting relevant texts and links into a notes app or email \cite{bota_self-es_2017}, taking screenshots and photos \cite{swearngin_scraps_2021}, or using a web clipper \cite{evernote_best_nodate}, collecting web content and encoding a user's mental context about it remains a cognitively and physically demanding process involving many different components: just the collection component itself involves deciding what and how much to collect, specifying the boundaries of the selection, copying it, switching context to the target application tab or window, transferring the information into the application where it will be stored \cite{faure_power_2009}, causing frequent interruptions to the users' main flow of reading and understanding the actual web content \cite{kittur_costs_2013,hinckley_informal_2012,marshall_saving_2005}, especially on mobile devices \cite{chang_supporting_2016,hahn_bento_2018}. In addition, components such as prioritizing options by importance result in additional overhead to move or mark their expected utility, which can change as users discover new options or old assumptions become obsolete. When further investigating each option, to keep track of evidence about its suitability, a user further needs to copy and paste each piece of evidence (e.g., text or images from a review or link to a video) and annotate it with how positive or negative it is relative to the user's goals.

Beyond the cognitive and physical overhead of collecting content and encoding context, prior work suggests that for learning and exploration tasks, people are often uncertain about which information will eventually turn out to be relevant and useful, especially at the early stages when there are many unknown unknowns \cite{bawden_perspectives_1999,fisher_distributed_2012,chang_supporting_2016,hahn_bento_2018}. This could further render people hesitant to exert effort to externalize their mental context if that effort might be later thrown away \cite{frederick_time_2002,liu_unakite:_2019}. One relevant example is Kittur's Clipper \cite{kittur_costs_2013,kittur_standing_2014}, which proactively prompted people to specify the ``valence'' (rating of good or bad) of an option (e.g., a specific camera) measured on a particular dimension (e.g., autofocus capability) as they collected information. Even though this elicitation of the mental model was done in situ and after much optimization of the interaction, it still required significant cognitive and physical effort and interruption, which prevented its widespread adoption. Other web clipping tools offer even less scaffolding for encoding mental context, typically only supporting a catch-all notes field that people rarely know how to take advantage of \cite{crescenzi_towards_2019}. 

To summarize, we frame a fundamental sensemaking challenge for people trying to research and make decisions online as the high friction involved in capturing: (1) the content that they want to keep track of, which can range from a word, a phrase, an image, to a paragraph or multiple blocks of mixed multimedia content, (2) which option or topic that content corresponds to and its perceived priority for further investigation (which is called ``triaging''), and (3) whether the evidence they find about that option or topic is positive or negative regarding its suitability for the user's goals (which is called ``valence'') \cite{liu_unakite:_2019}.\looseness=-1

Our vision in this work is to create a technique that reduces the friction for the transfer of a user's internal mental judgements while they are processing information into an external system that will capture those judgements and scaffold sensemaking and exploration. While it is a challenge for the cost of this transfer to be zero, we aim to reduce the overhead significantly by exploring a new class of gestures for this purpose based on ``wiggling:'' rapid back-and-forth movements of a pointer that do not require any clicking to perform, yet are sufficiently precise to accurately select the desired content, while at the same time supporting the optional and natural encoding of valence rating (positive to negative) and classification of priority (to facilitate triage) to the collected information (see Figure \ref{fig:wiggle-detailed-graph}). In addition, this technique does not conflict with typical existing interactions (like selecting text or clicking on hyperlinks) and can be extended to other device form factors such as touchscreens and mobile. The rating and classification can be applied by ending the ``wiggle'' with a swipe in different directions (see Figure \ref{fig:wiggle-detailed-graph}b,c,e).

We instantiate this class of wiggle-based gesture in an event-driven \javascript library and a prototype system called \systemname\footnote{\systemname stands for \systemnameInFull.}, which builds on top of an existing information and task management application called \basesystemname that already supports \textit{clipping} and assigning \textit{valence} to general web content as well as organizing them into \textit{topics} with \textit{priorities}. \systemname consists of a Chrome extension and a mobile application, which enables users to capture and classify information fluidly while searching and browsing. To combat the issue of potentially collecting too much information, the system enables users to easily filter and sort the collected information based on the encoding that the users applied at collection time (or later).\looseness=-1

In a lab evaluation with participants, we found that using \systemname to collect and triage information incurs 58\% less overhead cost to perform without sacrificing operational accuracy. In addition, participants generally preferred the wiggling techniques over \basesystemname alone due to its easiness and naturalness to perform as well as its ability to encode their mental contexts in an organic way.

The primary contributions described in this paper include:
\begin{itemize}[leftmargin=2em]
	\item A novel class of wiggle-based gestures that are cognitively and physically lightweight to perform to collect information, and can simultaneously encode aspects of users' mental context,
	\item A prototype event-driven \javascript library that implements such gestures and runs in web browsers,
	\item \systemname, a prototype system that takes advantage of the wiggle-based gestures to enable information capturing and classification during sensemaking that works on both desktop and mobile devices,
	\item A lab evaluation that offers empirical insights into the usability, usefulness, and effectiveness of the \systemname system.
\end{itemize}


\section{Related Work}

\subsection{Making Sense of Online Information}
This work builds on theories of sensemaking as defined as developing a mental model of an information space in service of a user's goals \cite{dervin_overview_1983,klein_making_2006,russell_cost_1993}. At a high level, the sensemaking process involves alternating between two phases: \textit{foraging}, which involves people searching for and extracting information, often from various data sources; and \textit{sensemaking}, the process of integrating the amassed information to form a schema or representation to interpret the space \cite{pirolli_sensemaking_2005}. In the following two sections, we present a brief review of prior studies and tools that are pertinent to these two phases.

\subsection{Capturing Information}
It has been reported that the foraging phase is where people spend the majority of time during a sensemaking process \cite{pirolli_information_1999,marchionini_information_1995,cepeda_distributed_2006,brandt_two_2009}. Therefore, there have been many research and commercial tools to try to help people better capture information during this phase. Some focused on keeping track of entire webpages or documents, such as SenseMaker \cite{baldonado_sensemaker:_1997} and browser bookmarks and reading lists; while others enabled users to capture finer-grain units within a web document, such as Hunter Gather \cite{schraefel_hunter_2002}, Clipper \cite{kittur_costs_2013}, and Google Notebook \cite{google_google_2012}. However, there is usually a high cost associated with these clipping mechanisms, from carefully maneuvering the cursor to specify the selection boundaries to frequent context switches between the content being read and the note-taking applications.

Prior work has also explored various ways to speed up the collection process. On the one hand, multiple approaches have been proposed to make \textit{selecting} desired content faster by offering pre-defined selection boundaries. For example, systems like Entity Quick Click and Citrine \cite{bier_entity_2006,ives_interactive_2009,stylos_citrine_2004} employ techniques like named-entity recognition \cite{mansouri_named_2008} to pre-process and highlight semantically meaningful entities in a document and allow users to collect and annotate relevant information with a single click. However, no matter how subtle the highlighting is, this could lead to distractions to a user's cognition process while comprehending the actual web content. \systemname, in contrast, leverages the organic boundaries that are readily defined on web content (e.g., individual words, phrases, as well as block level \html tags, such as \texttt{<div></div>}, \texttt{<p></p>}, \texttt{<img />}, etc.) \cite{chang_supporting_2016}, and only shows the boundaries when a wiggle is activated, which minimizes distractions to the reading experience.

On the other hand, research and commercial products have explored \textit{collecting} information on behalf of users as they search and browse the web. The history view in most modern browsers offers a linear depiction of a user's activities on a page level. Unfortunately, this format proves to provide little cue for helping people recall the information they have seen \cite{won_contextual_2009,leetiernan_two_2003}, and few people reported using the history feature \cite{weinreich_off_2006,aula_information_2005,jones_once_2002,jones_keeping_2001,byrne_tangled_1999}. Works such as Thresher \cite{hogue_thresher_2005} and Dontcheva et al.'s web summarization tool \cite{dontcheva_summarizing_2006} let users create and curate patterns and templates of information that they want to collect through examples, and then automatically collect that information from pages that users visit in the future. However, sensemaking is a dynamic process \cite{pirolli_sensemaking_2005}, especially during the foraging phase \cite{kittur_costs_2013}, and, unlike \systemname, these template-based approaches lack the necessary flexibility for users to capture whatever they want on the fly. Our recent system called Crystalline \cite{liu_crystalline_2022} explored having a system automatically collect potentially important information by leveraging natural language processing (NLP) techniques to understand the content users are browsing as well as signals from their browsing behavior, such as cursor positions \cite{huang_improving_2012}, clicks \cite{hijikata_implicit_2004}, and dwell time on a page \cite{claypool_implicit_2001} to implicitly indicate the user's interest. However, Crystalline is specifically designed and tuned for the domain of programming where web content is usually regularly positioned and formatted  \cite{hsieh_exploratory_2018,liu_unakite:_2019}. It is unclear how this approach would scale to general web content.

\subsection{Rating and Organizing Information}

Prior work has introduced various ways to incorporate information classification into the foraging phase. For example, Clipper \cite{kittur_costs_2013}, Unakite \cite{liu_unakite:_2019}, and Adamite \cite{horvath_understanding_2022} all prompt the user to optionally categorize an information clip after it has just been captured. Spar.tag.us \cite{hong_spartagus_2008} enables users to associate custom tags with individual paragraphs. ForSense \cite{rachatasumrit_forsense_2021} leverages natural language processing to automatically cluster information clips based on themes and topics. \systemname draws from and builds upon this prior work while focusing on selecting, clipping, rating,  and classifying the information all in one gesture.

There have also been a number of research tools developed to support in-depth organizing and structuring, such as the WebBook and WebForager by Card et al. \cite{card_webbook_1996}, which use a book metaphor to find, collect, and manage web pages and information, Webcutter, which collects and presents URL collections in tree, star, and fisheye views \cite{maarek_webcutter:_1997}, SenseMaker \cite{baldonado_sensemaker:_1997} for evolving collections of information, and Mesh \cite{chang_mesh_2020}, and our Unakite \cite{liu_unakite:_2019,liu_reuse_2021} which build comparison tables of various options and criteria. However, in the current work, we focus on helping users organize information into topics, a quick yet sufficiently expressive and more commonly-used structure \cite{crescenzi_towards_2019}, especially during the early stages of sensemaking.

\begin{figure*}[t]
\centering
\vspace{-2mm}
	\includegraphics[width=1.01\linewidth]{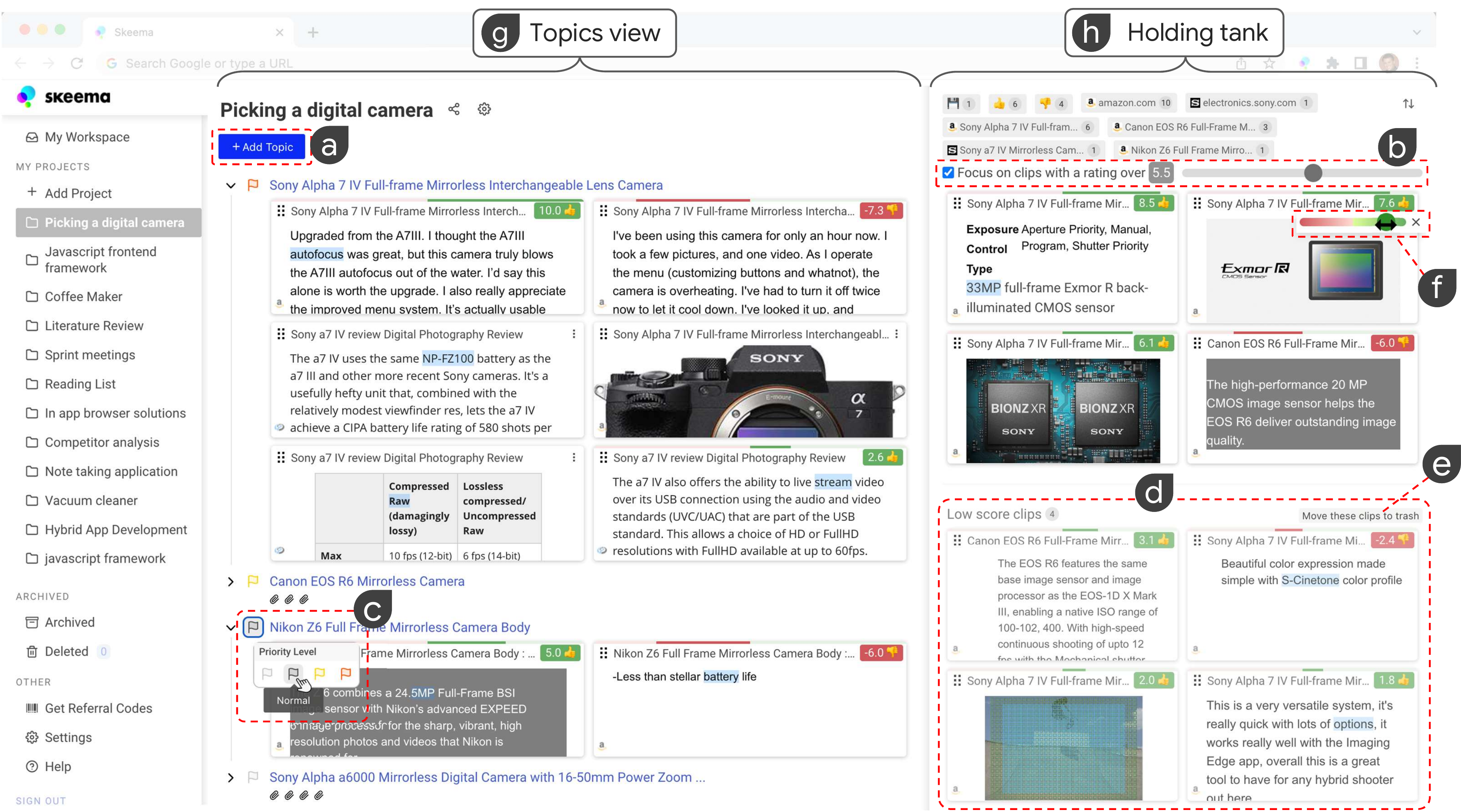}
	\vspace{-7mm}
	\caption{\systemname's UI built on top of \basesystemname. On the left is the topics view (g) where users can create a topic (a) as well as change its perceived priority (c). On the right is the holding tank (h) that holds the collected information, in which users can filter out information with a lower rating using the slider (b). As a result, clips with rating scores lower than the set threshold would be automatically grouped together at the end and grayed out (d), and users can easily archive or put them in trash by clicking a button (e). In addition, users can manually adjust the valence rating of an information clip (f).\looseness=-1}
	\label{fig:skeema-workspace-view}
	\vspace{-3mm}
\end{figure*}

\subsection{Recognizing and Using Gestures}
The wiggle gesture we use in \systemname (as shown in Figure \ref{fig:wiggle-detailed-graph}) has a similar form to a scratch-out gesture in some previous systems used for undo \cite{wobbrock_user-defined_2009}, edit \cite{rubine_specifying_1991}, or delete. Wiggling has also been used by some window managers, for example, Microsoft Windows 7 in 2009 introduced ``Aero Shake'' \cite{thurrott_windows_2009} where grabbing the title bar with the mouse and shaking the window left and right minimizes all other windows, or restores them. However, these gestures all require that the mouse button first be depressed, while our approach, on the contrary, is specifically designed to work when with none of the mouse buttons are depressed. In addition, macOS has an accessibility feature that supports shaking the mouse to make the size of the pointer much larger to help locate the pointer \cite{apple_support_make_2022}. Importantly, our testing shows that those features do not interfere with our browser-based implementation of wiggle-based gestures.

Over the years, many complex gesture \textit{recognizers} have been developed, such as the Rubine recognizer \cite{rubine_specifying_1991}, which extracts multiple features from a trajectory and uses a linear classifier for recognition. However, these parametric recognizers are difficult to control with respect to the variances in gestures to be supported. Another approach is template-based gesture recognition, such as the \$1 recognizer \cite{wobbrock_gestures_2007} and the Protractor recognizer \cite{li_protractor_2010}, which compare new trajectories to the pre-defined gesture templates, and is more lightweight without sacrificing too much accuracy. However, these recognizers can be both time and resource intensive, especially on mobile devices where the computing power and resources are usually limited. In our work, we built a heuristics-based ad-hoc recognizer (see section \ref{sec:system}), allowing the system to perform real-time eager recognition \cite{rubine_combining_1992} without impacting the performance of other UI activities on both desktop and mobile devices. In addition, building on prior evidence that people can accurately perform swipes to as many as eight different directions \cite{callahan_empirical_1988,kurtenbach_limits_1993}, we support ending the wiggle gesture with a directional swipe to further classify the collected information or encode people's mental context in situ.




\section{Background and Design Goals}

In this work, we explore using ``wiggling'' to select, clip, classify, as well as rate a piece of content in a single gesture during sensemaking, minimizing the interruption to people's main activities of reading and comprehending the content. To ground our research, we build on an existing information and task management system called \basesystemname. First, we briefly describe \basesystemname and its features related to the context of this work. Then, we discuss the design goals and processes for the wiggling gesture for the new \systemname system.\looseness=-1

\subsection{The \basesystemname system}
\basesystemname is a Chrome browser extension  designed to support people's need to collect and organize information and manage their tabs during online sensemaking. Different from general web clippers that typically only support saving entire pages of web content into an individual note within a notebook \cite{evernote_best_nodate}, \basesystemname enables people to save an arbitrary amount of web content as information clips (Figure \ref{fig:holding-tank-desktop-mobile}c) into a holding tank (Figure \ref{fig:skeema-workspace-view}h), and later organize them into topics in the topics view (Figure \ref{fig:skeema-workspace-view}g). For clipping, \basesystemname offers two methods: 

\begin{itemize}[leftmargin=1.2em]
    \item \textbf{Clipping text}: Users can select any arbitrary content in the usual way using the cursor and click the clipping button that pops up to collect the selected texts (see the upper part of Figure \ref{fig:skeema-basic-clipping} in the appendix).
    \item \textbf{Clipping screenshot}: Users can use the screenshot feature to drag out a bounding box to save the desired content (see the lower part of Figure \ref{fig:skeema-basic-clipping} in the appendix).
\end{itemize}


To help users express whether a piece of evidence that they collected is positive or negative with regard to their own goal, \basesystemname allows users to add a valence rating from -10 to +10, with negative values indicating a ``con'' and denoted by a ``thumbs-down'' emoji and positive values indicating a ``pro'' and denoted by a ``thumbs-up'' emoji (see Figure \ref{fig:holding-tank-desktop-mobile}c1). 

\basesystemname allows users to organize information into thematically related topics in the topics view (Figure \ref{fig:skeema-workspace-view}g). To achieve that, users need to manually create a topic (Figure \ref{fig:skeema-workspace-view}a), enter a name, and drag the desired information cards from the holding tank and drop it into the topic. Users can also set priority to a topic to indicate its perceived utility and how much they want to follow up on it, which defaults to be ``Normal'', but can also be set to ``Low'', ``High'', or ``Very high'' (Figure \ref{fig:skeema-workspace-view}c).

Although \basesystemname has the support for collecting finer-grain content (which research has shown to be the unit of information that people usually think in and work with during sensemaking \cite{schraefel_interaction_2001,marshall_saving_2005}), there is still a high cost in specifying the collection boundary and adding ratings and priorities to the collected information and topics (which users would have to switch to the \basesystemname tab to do). In addition, clipping text in \basesystemname loses the text's original \css styling, which might be helpful for quicker recognition later on \cite{liu_unakite:_2019}, and \basesystemname does not gracefully support collecting consecutive blocks of mixed content (e.g., consumer review text of a camera followed up some sample photos, such as shown in Figure \ref{fig:skeema-basic-clipping}b).

\subsection{Design Goals for Low-cost Information Capturing and Triaging}\label{sec:design-goals}
Guided by prior work and well as the limitations of \basesystemname discussed above, we set out to provide an interaction that could simultaneously reduce the cognitive and physical costs of \textit{capturing} information while providing natural extensions to easily and optionally \textit{encode} aspects of users' mental context during sensemaking. We hypothesize that such an effective interaction should have the following characteristics:

\begin{enumerate}[label=(\arabic{*}),leftmargin=1.5em,labelsep=*]
	\item \textbf{Accuracy}: It needs to be accurate and precise enough to lock onto the content the users intend to collect.\label{dg:accuracy}
	
	\item \textbf{Efficiency}: It should be quick and low-effort to perform, and minimize interruptions to the main activities that users are performing, such as learning and active reading.\label{dg:efficiency}
	
	\item \textbf{Expressiveness}: It should be extendable to provide natural and intuitive affordances for users to express aspects of their mental context at the moment. In the scope of this work, we would like to have wiggling support encoding valence ratings as well as topic priorities. \label{dg:expressiveness}
	
	\item \textbf{Integration}: It should be a complement to and not interfere with the existing interactions that users already use, such as using the pointer to select text and pictures or click on links.\label{dg:integration}
\end{enumerate}

Below, we present a brief overview of the iterative design exploration leading to the current wiggle-based interactions.




\subsection{Iterative Design Exploration}

To begin our exploration, we took a desktop-first approach and brainstormed various interactions that would address these four design goals. To ground our explorations, we also prototyped these candidate interactions using \javascript in a browser, which is where a large portion of the reading and collecting happens \cite{hahn_bento_2018,chang_supporting_2016}. Like previous approaches, collecting the desired content, including text and/or images, can be broken down into two main phases: \textit{(a) identifying the desired target} and \textit{(b) triggering the collection}.

One of the interactions we first explored was simply clicking on the desired content (or in the gutter to the left or right) to capture it into the system, similar to existing interactions supported by some text editors such as Microsoft Word. Although straightforward, this interferes with existing selection methods, and would require users to first enter a ``grabber'' mode, possibly through a special hotkey combination, which violates both design goals \ref{dg:efficiency} and \ref{dg:integration}. Next, we experimented with hovering the pointer over the target content and keeping it still for a period of time in order to trigger a collection. This has the benefit of not interfering with existing interaction methods as there is no clicking required, satisfying goal \ref{dg:integration}. However, research has shown that when heavily engaged in active reading and sensemaking tasks, people often need to select and save information frequently within short time intervals \cite{chang_supporting_2016,whittaker_personal_2011}, and waiting for a noticeable amount of time will add an inherent cost to every collection operation a user wants to perform and therefore is likely to interrupt the user's main activity, violating design goal \ref{dg:efficiency}.\looseness=-1

Next, we experimented with using non-click gestures (satisfying goal \ref{dg:integration}) performed on the desired target to trigger the selection, since gestures are considered intuitive to perform and widely used in both commercial and academic systems \cite{rubine_combining_1992,vatavu_stroke-gesture_2019,le_shortcut_2020,lank_sloppy_2005,li_braillesketch_2017}. One of the promising ideas was to use the mouse pointer to sketch out a certain shape over the desired target to trigger a collection. In addition, by varying the shape, it could theoretically support encoding different aspects of users' mental model, such as sketching a ``+'' for marking it as a ``pro'' and ``-'' as a ``con'' \cite{wobbrock_gestures_2007}, supporting design goal \ref{dg:expressiveness}. However, similar to using keyboard shortcuts, it is hard for users to learn and memorize the different shapes without special affordances \cite{zheng_fingerarc_2018,le_shortcut_2020}. Furthermore, making sure one sketches out the correct gesture may require non-trivial physical as well as cognitive demand, violating goal \ref{dg:efficiency}, and even so, these shapes can have a high false recognition rate, violating goal \ref{dg:accuracy}.

We then experimented with gestures that do not require special training or practice in order to perform accurately. One that worked particularly well is wiggling the mouse pointer, i.e., making small ballistic back-and-forth movements, on top of the desired collection target (Figure \ref{fig:wiggle-detailed-graph}a). Here, the choice of a target could be determined from the average or starting location of the mouse pointer during the gesture, and the user continues to perform the same back-and-forth motion until reaching a certain threshold to trigger the collection. Indeed, prior work has suggested that people naturally use the mouse pointer to guide their attention while reading \cite{hinckley_informal_2012}, or even unconsciously have the pointer follow their eye gaze \cite{huang_improving_2012}, so the pointer could be readily available to initiate a wiggle in place.\looseness=-1

This has some additional benefits, such as it seemed natural and intuitive like scratching off something \cite{wobbrock_user-defined_2009,rubine_specifying_1991}, it can be activated without clicking, which can be both cognitively and physically costly \cite{kittur_costs_2013}, and is robust against false positives since only a very specific motion pattern could trigger a collection. Furthermore, it can be chained with optional operations such as swiping in different directions that not only are consistent with the wiggling gesture itself but also intuitively map to users' mental context (such as swiping left/right for negative/positive and up/down for various levels of importance, and even leveraging the amount of distance traveled of a swipe to encode a continuous value).


Since there is no mouse pointer on mobile devices such as smartphones, and using fingers to move left and right in browsers triggers page navigation back and forth, whereas up and down is used for scrolling, we decided to take advantage of these small up-and-down scroll events, since they are not currently in use by any existing interactions. Therefore the wiggling counterpart on mobile devices became using the finger to quickly scroll up-and-down while the finger is over the desired collection target (Figure \ref{fig:wiggle-detailed-graph}d). 

\begin{figure*}[t]
\centering
\vspace{-3mm}
	\includegraphics[width=1.00\linewidth]{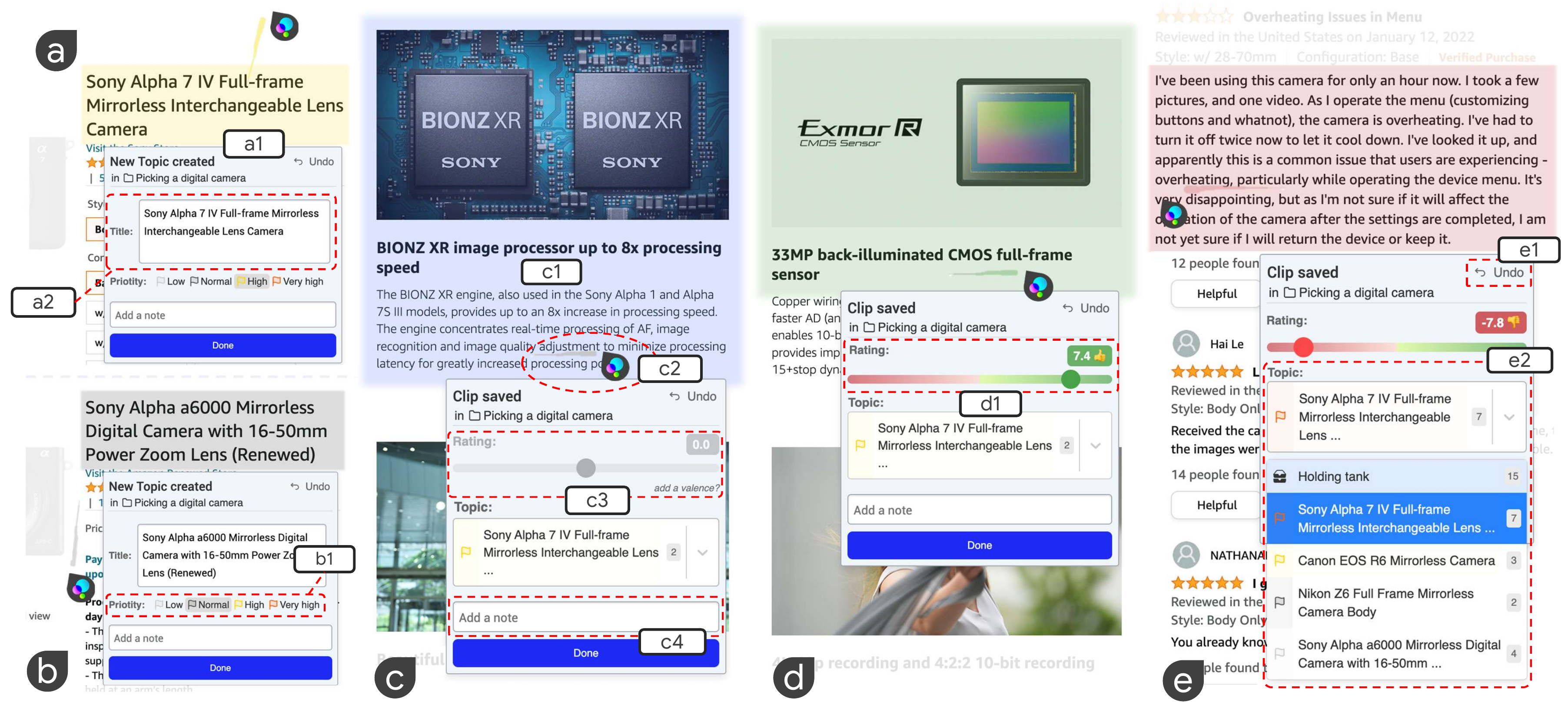}
  \vspace{-7mm}
  \caption{Using wiggling to collect information as well as encode priorities and valence ratings. Specifically, as shown in (c), users can wiggle (c2) over the desired content (c1) to collect it into the information holding tank (Figure \ref{fig:holding-tank-desktop-mobile}c). A popup dialog will be presented near the just collected content to allow users to optionally add a valence rating (c3), pick a topic that the content should go into (e2), add notes (c4), as well as undo the collection (e1). In addition to regular collection, users can also end the wiggle with a swipe right to encode a positive rating (d) or left to encode a negative rating (e), which can also be changed in the popup dialog (d1). Furthermore, by ending a wiggle with a swipe up (a) or down (b), users can create a new topic with different priorities (b1), and can change the title of the topic directly in the popup dialog (a1).}
  \label{fig:popup-views-crop}
  \vspace{-3mm}
\end{figure*}

\section{The \systemname System}\label{sec:system}

\subsection{Wiggle-based Gestures}\label{sec:wiggle-gestures-formalization}
For desktop computers with a traditional mouse, trackpad or trackball input device, the wiggle interaction consists of the following stages, as illustrated in Figure \ref{fig:wiggle-detailed-graph}a,b,c:

\begin{enumerate}[label=(\arabic{*}),leftmargin=1.5em,labelsep=*,topsep=3pt]
    \item \textbf{Acquiring the collection target}: To initiate, users move their mouse pointer onto the target content that they would like to collect (Figure \ref{fig:wiggle-detailed-graph}a0) and initiate the wiggling movement specified in the steps below. \systemname uses an always-on wiggle gesture recognizer to automatically detect the start of a wiggling gesture. This avoids the requirement of an explicit signal like a keyboard key or mouse down event, which might conflict with other actions, and has the benefit of combining activating and performing the gesture together into a single step, therefore reducing the starting cost of using the interaction technique.
    
    \item \textbf{Wiggle}: To collect the target content, users simply move the mouse pointer left and right approximately inside the target content. To indicate that the system is looking to detect the wiggling gesture, it will display a small ``tail'' (e.g., Figure \ref{fig:popup-views-crop}c2) that follows the pointer on the screen, and replaces the regular pointer with a special one containing the \basesystemname icon. \systemname also adds a dotted blue border to the target content to provide feedback about what content will be collected, and the blue color grows in shade as users perform more lateral mouse movements (Figure \ref{fig:wiggle-detailed-graph}a1-4). This is analogous to half-pressing the shutter button to engage the auto-focus system to lock onto a subject when taking photos with a camera. To assist with collecting fine grain targets, ranging from a word to a block (e.g., a paragraph, an image), \systemname allows users to vary the average size of their wiggling to indicate the target that they would like to collect: if the average size of the last five lateral movements of a pointer is less than 65 pixels (a threshold empirically tuned that worked well in our pilot testing and user study, but implemented as a customizable parameter that individuals can tune based on their situations), \systemname will select the word that is covered at the center of the wiggling paths; while larger lateral movements will select a block-level content (details discussed in section \ref{sec:consideration-targeting}). In addition, users can abort the collection process by simply stopping wiggling the mouse pointer before there are sufficient back-and-forth movements.\looseness=-1
    
    \item \textbf{Collection}: As soon as users make at least five back and forth motions (optimized for the amount of physical effort required and the number of false positive detection through pilot testing, but is also implemented as a parameter that can be customized by individuals in practice, details discussed in section \ref{sec:consideration-recognition}), the system will commit to the collection, and gives the target a darker blue background showing that a wiggle has been successfully activated (as shown in Figure \ref{fig:wiggle-detailed-graph}a5). If users want to collect multiple blocks of content, they can just naturally continue to wiggle over other desired content after this activation. Or, they can stop wiggling. However, if users have selected the wrong target, an undo button appears, which can be clicked to cancel the collection (Figure \ref{fig:popup-views-crop}e1).\looseness=-1

    \item \textbf{Extension}: Instead of just stopping the wiggle motion after collection, users can leverage the last wiggle movement and turn it into a ``swipe'', either horizontally to the right or left to encode a positive or negative valence rating (as shown in Figure \ref{fig:wiggle-detailed-graph}b1,b2), or vertically down or up to specify a topic and priority for that topic (as shown in Figure \ref{fig:wiggle-detailed-graph}c1-4). Feedback for the extension uses different colors for the background of the target content to provide visual salience (details discussed in section \ref{sec:system-overview}).\looseness=-1
\end{enumerate}

\noindent Similarly, on a mobile device with touch screens:

\begin{enumerate}[label=(\arabic{*}),leftmargin=1.5em,labelsep=*,topsep=3pt]
    \item \textbf{Acquiring collection target}: To initiate, a user's finger touches the target content that should be selected. 
    
    \item \textbf{Wiggle}: To collect the target block, the user keeps the finger on the screen and starts making small up-and-down scrolling movements. Similar to the desktop scenario, the system adds a dotted blue border to the target content to provide feedback that the wiggling is being detected (Figure \ref{fig:wiggle-detailed-graph}d0-4). Note that due to the limitations of the large size of the finger with respect to an individual word \cite{chang_supporting_2016} as well as the unique use cases of mobile devices (e.g., quickly consuming and collecting blocks of information on the go \cite{iqbal_multitasking_2018,williams_mercury_2019}), \systemname for mobile only supports selecting block-level content such as paragraphs or images.\looseness=-1
    
    \item \textbf{Collection}: As soon as the user makes at least five up-and-down motions, the system will commit to the collection by giving the target a darker blue background (Figure \ref{fig:wiggle-detailed-graph}d5). Now, the user can stop wiggling and lift the finger from the screen. Similar to the desktop version, an undo button pops up that lets the user cancel the collection in case of an error. Note that due to the limited screen real estate that typical mobile devices afford, additional blocks of content will have to be first scrolled into view for users to then capture them, which would make the interaction less fluid. Therefore, collecting multiple blocks of content is currently not supported by \systemname on mobile.\looseness=-1

    \item \textbf{Extension}: Instead of stopping the wiggle motion after collection, users can end the wiggle with a horizontal swipe to the left or right to achieve similar encoding capabilities described for the desktop version. After the system detects the wiggle, it turns off other actions until the finger is lifted, so the swipes do not perform their normal actions. (But the normal swipes, scrolling, and other interactions still work normally when not preceded by a wiggle.) Currently, since \systemname already uses the vertical dimension for detecting wiggling movement on a mobile device, and large cross-screen vertical movements are difficult to perform, especially when holding and interacting with a single hand, we opted not to make a mobile equivalent of encoding topic priorities.\looseness=-1
\end{enumerate}

\subsection{An Overview of The \systemname System}\label{sec:system-overview}
\systemname enables users to collect and triage web content via wiggling. First of all, after a regular wiggle with no extension (Figure \ref{fig:popup-views-crop}c), \systemname presents a popup dialog (augmenting the original \basesystemname popup) directly near the collected content to indicate success. In addition to \basesystemname's notes field (Figure \ref{fig:popup-views-crop}c4), users can attach a valence rating (Figure \ref{fig:popup-views-crop}c3) and pick the topic that this piece of information should be organized in (Figure \ref{fig:popup-views-crop}e2), as opposed to post-hoc organization using drag and drop as required by \basesystemname. By default, it goes into the last topic the user picked or the holding tank if none was picked initially. 
Unlike \basesystemname where information was saved in pure text format or an inflexible screenshot with limited resolution, \systemname leverages the technique introduced in \cite{liu_unakite:_2019,liu_crystalline_2022} to preserve and subsequently show the content with its original \css styling, including the rich, interactive multimedia objects supported by HTML, like links and images. This makes the content more understandable and useful, and also helps users quickly recognize a particular piece of information among many others by its appearance \cite{liu_unakite:_2019}.

Of course, a more fluid way to encode user judgements than what was described above is to leverage the natural extension of the wiggle gesture discussed in the previous section: to encode a valence rating in addition to collecting a piece of content, users can end a wiggle with a horizontal ``swipe'', either to the right to indicate positive rating (or ``pro'', characterized by a green-ish color that the background of the target content turns into, and a thumbs-up icon, as shown in Figure \ref{fig:popup-views-crop}d), or the left for negative rating (or ``con'', characterized by a red-ish color that the background of the collected block turns into, and a thumbs-down icon, as shown in Figure \ref{fig:popup-views-crop}e). Optionally, users can also turn on real-time visualizations of ``how much'' they swiped to the left or right to encode a rating score representing the degree of positivity or negativity, and can adjust that value in the popup dialog (Figure \ref{fig:popup-views-crop}d1) or from the information card (Figure \ref{fig:skeema-workspace-view}f). Under the hood, \systemname calculates this score as the horizontal distance the pointer traveled leftward or rightward from the average wiggle center divided by the available distance the pointer could theoretically travel until it reaches either edge of the browser window. This score is then scaled to be in the range of -10 to 10 to match with the existing values provided by \basesystemname. 

Alternatively, to directly create a topic and encode it with a priority from wiggling, users can either end the wiggle with a swipe up (encoding ``high'', characterized by a yellow-ish color that the background of the target content turns into, as shown in Figure \ref{fig:popup-views-crop}a) or down (encoding ``normal'', characterized by a gray-ish color that the background of the target content turns into, as shown in Figure \ref{fig:popup-views-crop}b). Optionally, if the user swipes all the way up or down to the edge of the browser window, \systemname will additionally encode two more levels of priorities, ``urgent'' and ``low'', indicated by a bright orange and a muted gray color (Figure \ref{fig:popup-views-crop}b1), which can be adjusted in the popup dialog (Figure \ref{fig:popup-views-crop}b1) as well as in the topics view (Figure \ref{fig:skeema-workspace-view}c). In this case, the content will instead be used as the default \textit{title} of the newly created topic (which users can change in the popup dialog directly as shown in Figure \ref{fig:popup-views-crop}a2 or later in the topics view).\looseness=-1

\begin{figure}[t]
\centering
	\includegraphics[width=1\linewidth]{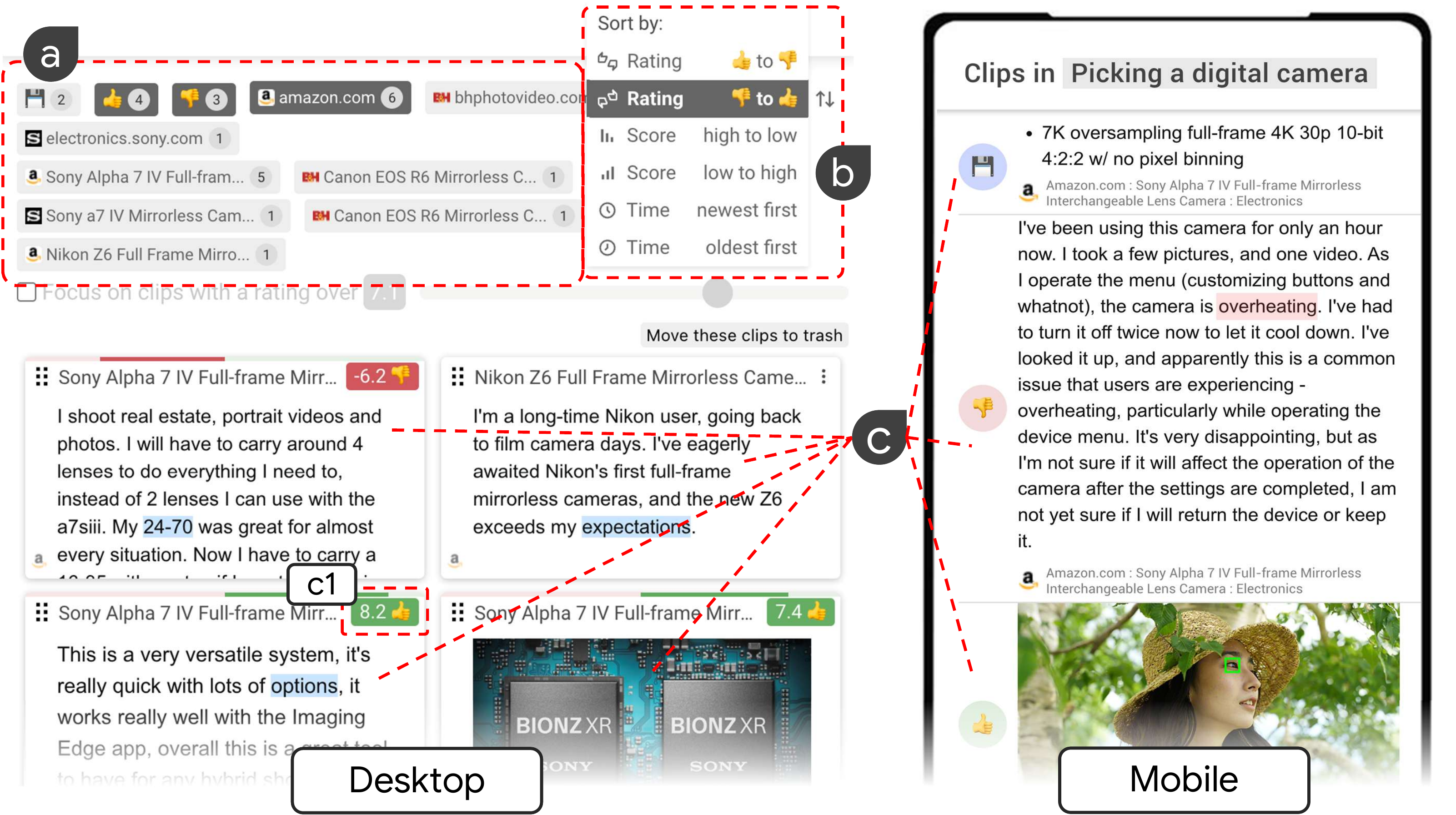}
	\vspace{-7mm}
	\caption{\systemname's information holding tank shown both on desktop and on mobile, which houses content that users collected through wiggling in the form of information cards (c). In addition, on desktop, users can apply different filters (a) and sorting mechanisms (b) to the information cards.}
	\label{fig:holding-tank-desktop-mobile}
	\vspace{-5mm}
\end{figure}

To help users better manage the information that they have gathered in the holding tank, \systemname offers several additional features on top of the original \basesystemname system. First, it enables users to sort the information cards by various criteria, such as in the order of valence ratings or in temporal order (Figure \ref{fig:holding-tank-desktop-mobile}b). Second, it offers category filters (Figure \ref{fig:holding-tank-desktop-mobile}a) automatically generated based on the encodings that users provided using wiggling (or edited later) and the provenance of information (where it was captured from). Users can quickly toggle those on or off to filter the collected information. For example, in Figure \ref{fig:holding-tank-desktop-mobile}b, the information with a ``positive rating'' or ``negative rating'' and collected from ``amazon.com'' was filtered and shown, as indicated by the dark gray background of the corresponding filters (if none of the filters are enabled, all the information cards will be shown). Third, users can quickly filter out information with a lower rating (e.g., indicating that it was less impactful to a user's overall goal and decision making) by adjusting the threshold using the ``Focus on clips with a rating over \texttt{threshold}'' slider shown in Figure \ref{fig:skeema-workspace-view}f. As a result, clips with rating scores lower than the set threshold would be automatically grouped together at the end and grayed out (Figure \ref{fig:skeema-workspace-view}d), and users can easily archive or put them into the trash in a batch by clicking the ``Move these clips to trash'' button (Figure \ref{fig:skeema-workspace-view}e). These organizational features further help users reduce clutter in the holding tank, and provide a scaffold for them to start dragging and dropping clips into their respective topics.\looseness=-1

Due to the limited screen size and use cases of a mobile device, we chose to only let users view the clips along with their valence in the holding tank (Figure \ref{fig:holding-tank-desktop-mobile}c).

\subsection{Design and Implementation Considerations}
In this section, we discuss important design and implementation considerations made through prototyping \systemname with \javascript in a browser to achieve the design goals specified in section \ref{sec:design-goals}.

\subsubsection{Recognizing a wiggle gesture}\label{sec:consideration-recognition}
For accurately recognizing the wiggle pattern, we explored several options. One way is to use an off-the-shelf gesture recognizer such as the \$1 \cite{wobbrock_gestures_2007} or the Protractor \cite{li_protractor_2010} recognizer. Although these recognizers may be lightweight and easy to customize, they are fundamentally designed to recognize distinguishable shapes such as circles, arrows, or stars, while the path of our wiggle gesture does not conform to a particular shape that is easily recognizable (and we argue that it should not conform to any particular shape, the sketching of which would increase the cognitive and physical demand).
A second option we investigated was to build a custom computer vision based wiggle recognizer using transfer learning from lightweight image classification models such as MobileNets \cite{howard_mobilenets_2017}. Though these ML-based models improved the recognition accuracy in our internal testing, they incurred a noticeable amount of delay due to browser resource limitations (and limitations in network communication speed when hosted remotely).
This made it difficult for the system to perform eager recognition \cite{rubine_combining_1992} (recognizing the gesture as soon as it is unambiguous rather than waiting for the mouse to stop moving), which is needed to provide real-time feedback to the user on their progress.

To address these issues, we discovered that a common pattern in all of the wiggle paths that users generated with a mouse or trackpad during pilot testing share the characteristic that there were at least five (hence the activation threshold mentioned in section \ref{sec:wiggle-gestures-formalization}) distinguishable back and forth motions in the horizontal direction, but inconsistent vertical direction movements. Similarly, on smartphones, wiggling using a finger triggers at least five consecutive up and down scroll movements in the vertical direction but inconsistent horizontal direction movements. Therefore, we hypothesized that only leveraging motion data in the principle dimension (horizontal on desktop, and vertical on mobile) would be sufficient for a custom-built recognizer to differentiate intentional wiggles from other kinds of motions by a cursor or finger. 

Based on our implementation using \javascript in the browser, we found that it successfully supports real-time eager recognition with no noticeable impact on any other activities that a user performs in a browser. Specifically, the system starts logging all mouse movement coordinates (or scroll movement coordinates on mobile devices) as soon as any mouse (or scroll) movement is detected, but still passes the movement events through to the rest of the DOM tree elements so that regular behavior would still work in case there is no wiggle. In the meantime, the system checks to see if the number of reversal of directions in the movement data in the principle direction exceeds the activation threshold, in which case a ``wiggle'' will be registered by the system. After activation, the system will additionally look for a possible subsequent wide horizontal or vertical swipe movement (for creating topics with priority or encoding valence to the collected information) without passing those events through to avoid unintentional interactions with other UI elements on the screen.
As soon as the mouse stops moving, or the user aborts the wiggle motion before reaching the activation threshold, the system will clear the tracking data to prepare for the next possible wiggle event.

\subsubsection{Target Acquisition}\label{sec:consideration-targeting}


In order to correctly lock onto the desired content without ambiguity, we explored two approaches that we applied in concert in \systemname. The first approach is to constrain the system to only be able to select certain targets that are usually large enough to contain a wiggling path and semantically complete. For example, one could limit the system to only engage wiggle collections on block-level semantic elements \cite{noauthor_block-level_nodate}, such as \code{<div>}, \code{<p>}, \code{<h1>-<h6>}, \code{<li>}, \code{<img>}, \code{<table>}, etc. This way, the system will ignore inline elements that are usually nested within or between a block-level element. This approach, though sufficient in a prototype application, does rely on website authors to organize content with semantically appropriate \html tags.  

The second approach is to introduce a lightweight disambiguation algorithm that detects the target from the mouse pointer's motion data in case the previous one did not work, especially for a small \texttt{<span>} or an individual word. To achieve this, we chose to take advantage of the pointer path coordinates (both X and Y) in the last five lateral mouse pointer movements, and choose the target content covered by the most points on the path. Specifically, we used the same re-sampling and linear interpolation technique introduced in the \$1 gesture recognizer \cite{wobbrock_gestures_2007} to sample the points on a wiggle path to mitigate variances caused by different pointer movement speeds as well as the frequency at which a browser dispatches mouse movement events.

On mobile devices, since the vertical wiggling gesture triggers the browser's scrolling events, the target moves with and stays underneath the finger at all times. Therefore, we simply find the target under the initial touch position. 

When \systemname is unable to find a target (e.g., when there is no HTML element underneath where the mouse pointer or the finger resides) using the methods described above, it does not trigger a wiggle activation (and also not the aforementioned set of visualizations), even if a ``wiggle action'' was detected. This was an intentional design choice to further avoid false positives as well as to minimize the chances of causing distractions to the user.

\subsubsection{Integration with existing interactions}\label{sec:consideration-integration}
Notice that the wiggling interaction does not interfere with common active reading interactions, such as moving the mouse pointer around to guide attention, regular vertical scrolling or horizontal swiping (which are mapped to backward and forward actions in both Android and iOS browsers) \cite{tashman_active_2011, morris_reading_2007}. In addition, wiggling can co-exist with conventional precise content selection that are initiated with mouse clicks or press-and-drag-and-release on desktops or long taps or edge taps on mobile devices \cite{chen_bezelcopy_2014,roth_bezel_2009}. Furthermore, unlike prior work that leverages pressure-sensitive touch screens to activate a special selection mode \cite{chang_supporting_2016}, the wiggling interaction does not require special hardware support, and can work with any kind of pointing device or touch screen.

\subsection{Implementation Notes}

We implemented the wiggling technique as an event-driven Java-Script library that can be easily integrated into any website and browser extension. Once imported, the library will dispatch \texttt{wiggle}-related events once it detects them. Developers can then subscribe to these events in the applications that they are developing. All the styles mentioned above are designed to be easily adjusted through predefined \css classes. The library itself is written in approximately 1,100 lines of \javascript and \typescript code.

The \systemname browser extension is implemented in \html, \typescript, and \css and uses the React \javascript library \cite{facebook_react_2018} for building UI components. It uses Google Firebase for backend functions, database, and user authentication. In addition, the extension is implemented using the now standardized Web Extensions APIs \cite{mozilla_browser_2022} so that it would work on all major browsers, including Google Chrome, Microsoft Edge, Mozilla Firefox, Apple Safari, etc. However, we primarily targeted Google Chrome and Microsoft Edge to minimize testing efforts during development.

The \systemname mobile application is implemented using the Angular \javascript library \cite{google_angular_2019}, the Ionic Framework \cite{ionic_cross-platform_nodate} and works on both iOS and Android  operating systems. Due to the limitations that none of the current major mobile browsers have the necessary support for developing extensions, \systemname implements its own browser using the \texttt{InAppBrowser} plugin from the open-source Apache Cordova platform \cite{apache_apache_nodate} to inject into webpages the \javascript library that implements wiggling as well as custom \javascript code for logging and communicating with the Firebase backend.\looseness=-1

\section{User Evaluation}
We conducted an initial lab study to evaluate the usability and usefulness of \systemname in helping people collect information as well as encode aspects of their mental context while doing so. Specifically, we aimed to address the following research questions:

\begin{itemize}[leftmargin=2em]
	\item \textbf{RQ1 [Accuracy]}: Are wiggle-based interactions sufficiently accurate to help users collect what they want?
	\item \textbf{RQ2 [Efficiency]}: Are wiggle-based interactions sufficiently low-friction to perform without interrupting the primary reading and sensemaking activities?
	\item \textbf{RQ3 [Expressiveness]}: Are the proposed extensions of marking priorities and valence useful in helping people encode their mental contexts?
	\item \textbf{RQ4 [Integration]}: Do wiggle-based interactions interfere with existing interactions that people are already using?
\end{itemize}

\subsection{Participants}
We recruited 12 participants (6 male, 6 female; 3 students, 3 software engineers, 2 UX designers, 1 UX researcher, 1 medical doctor, 1 administrative staff member, and 1 entrepreneur) aged 21-38 years old (mean age = 28.5, $SD$ = 4.5) through emails and social media. Participants were required to be 18 or older and fluent in English. All participants reported experience reading and making sense of large amounts of information online for either professional or personal purposes on a daily basis, and had tried or were using commercially available web clipping and organization tools and systems, such as the Evernote Clipper, OneNote, or Notion.

\subsection{Study Methodology}
The study was a within-subjects design with each participant engaging in two tasks, one using \systemname with \basesystemname in the experimental condition, and the other just using \basesystemname in the control condition, counterbalanced for order. For our control condition, \basesystemname provided the affordances of a web clipping tool, which would provide a more conservative and matched baseline than no tool support.
Specifically, our control condition enabled participants to capture text through a popup button (Figure \ref{fig:skeema-basic-clipping}a1) to save highlighted text and a screenshot clipper instead of the wiggle interaction. After saving the information, participants could set the priority of topics and the valence of information in the workspace view (Figure \ref{fig:skeema-workspace-view}), versus being able to encode them as a continuation of the wiggle in \systemname. 

For each task, participants were presented with a product category they needed to research, and a set of three Amazon pages from which they were required to collect information. Participants were instructed to read through the provided webpages, collect information, and organize the information clips into topics, such as by different options or different criteria in which the options should be evaluated. They were required to at least collect 10 information clips as well as create a minimum of 3 topics with priority for each task. Participants had 15 minutes to complete the task, but could inform the experimenter to move on if they finished early. 

The two tasks were:

\begin{itemize}[leftmargin=2em]
	\item \textit{(A)} Choosing a digital mirrorless camera: participants were told to imagine that they were to purchase a new mirrorless camera to take photos of their spouse and young kids on their weekend road trips.
	\item \textit{(B)} Buying a vacuum cleaner: participants were told to imagine that they were to buy a new vacuum cleaner in preparation for moving into a new house with a newborn baby and their two pets.
\end{itemize}

In order to minimize differences between tasks and participant decision making, we provided a fixed set of web pages per task, each with approximately eight screens of content. As described in the results, the two tasks took approximately the same amount of time for participants to finish, and were counterbalanced in order and randomized across conditions.

Each study session started by obtaining consent and having participants fill out a demographic survey. Participants were then given a 10-minute guided tutorial showcasing the various features of \systemname as well as the baseline system, and a 10-minute free-form practice session to familiarize themselves with the features of both systems. At the end of the study, participants completed a survey and engaged in a semi-structured interview about their experience with the tool. The interview focused on participants’ perceptions of using the wiggle-based interactions. The questions probed the perceived effectiveness of wiggling, their current practices around collecting information, and scenarios where they thought wiggling would be useful and how they would modify it to be more useful. The interviews were audio-recorded and transcribed, after which qualitative coding and thematic analysis \cite{charmaz_constructing_2006} were performed.

Each study session took approximately 60 minutes to complete, using a designated MacBook Pro computer with Google Chrome and \systemname installed as well as a Logitech MX Master 2S mouse. All sessions were carried out in person, with the participants and the experimenter appropriately masked following COVID-19 mitigation guidelines. All participants were compensated \$15 for their time. The study was approved by our institution's IRB.

\newcommand{\na}{N/A\xspace}
\newcommand{\statsSig}[1]{#1*}
\newlength{\rowpaddingbottom}
\setlength{\rowpaddingbottom}{1.2mm}
\begin{table*}[t]
\vspace{-2mm}
\centering
\resizebox{1\textwidth}{!}{%
\begin{tabular}{
c|
p{14mm}|
p{20mm}|
p{22mm}|
p{26mm}|
p{26mm}|
p{43mm}|
p{26mm}
}
\toprule
&
\textbf{Condition} & 
\textbf{Overhead cost}  &
\textbf{Time (seconds)}  &
\textbf{$n$ of clips collected}  & 
\textbf{$n$ of topics created using wiggling}  &
\textbf{$n$ of topics created separately in the workspace view}  &
\textbf{Total $n$ of topics created}   \\
\midrule
    
\multirow{2}{*}{{\textbf{Task A}}} & 
Baseline  &
33.0\% (8.60\%) &
713.7 (76.0) &
21.0 (7.03) &
\na &
4.17 (1.17) &
4.17 (1.17) 
\\
    
&
\systemname  &
14.0\% (7.89\%) &
558.7 (76.5) &
38.3 (5.28) &
7.50 (1.05) &
0.50 (0.84) &
8.00 (1.89) 
\\
\midrule

\multirow{2}{*}{{\textbf{Task B}}} & 
Baseline  &
30.40\% (7.31\%) &
692.0 (131.4) &
19.5 (6.81) &
\na &
4.67 (0.52) &
4.67 (0.52) 
\\

&
\systemname  &
12.8\% (2.74\%) &
515.7 (54.3) &
37.3 (8.64) &
7.17 (1.17) &
0.50 (1.22) &
7.67 (2.39) 
\\
\midrule

\multirow{2}{*}{{\textbf{Average}}} & 
Baseline  &
31.7\% (7.73\%) &
702.8 (102.9) &
20.3 (6.68) &
\na &
4.42 (0.90) &
4.42 (0.90) 
\\

&
\systemname  &
13.4\% (5.67\%) &
536.8 (67.0) &
37.8 (6.85) &
7.33 (1.07) &
0.50 (1.00) &
7.83 (2.07) 
\\
\bottomrule
\end{tabular}%
}
\vspace{0mm}
\caption{Statistics for various performance measures in the user study. Standard deviations are included in the parentheses.}
\label{tab:stats-table}
\vspace{-6mm}
\end{table*}

\section{Results}
All participants were able to complete each task within the specified 15 minute time limit. Below, we compile together both quantitative and qualitative evidence to evaluate \systemname with respect to our four design goals and research questions.

\subsection{RQ1 [Accuracy]}
First, evaluate if the wiggling gestures are accurate enough to help users collect and express what they want. Specifically, we looked for cases where: (1) participants hit the undo button to dismiss an incorrect wiggle activation and redo the wiggling due to \systemname picking up the wrong target content, which turned out to be on average 0.67 (SD = 0.65) times per person per task, and only accounted for 1.48\% of the 45.16 (SD = 8.82) total wiggle actions participants on average performed per task; (2) participants had to use the popup dialog to immediately edit the valence or topic priority because \systemname picked the wrong swipe direction, which turned out to be 0; (3) participants had to redo the wiggling gesture because the previous one they performed did not activate at all, which turned out to be on average 0.92 (SD = 0.67) times per person per task, and only accounted for 2.01\% of the total wiggle actions participants on average per task.

This evidence suggests that the wiggling technique provided by the current \systemname system is sufficiently accurate and robust, at least with ample amount of training and practice. It would be interesting for future work to explore how it performs in the wild, potentially without much upfront practice, and examine whether and how people's wiggling accuracy and performance evolve over time.\looseness=-1




\subsection{RQ2 [Efficiency]}
Second, we are interested in understanding if \systemname creates a more fluid experience when collecting and triaging information with less interruption compared to the baseline condition. For this comparison, we opted to measure two key metrics: the \textit{overhead cost} of using a tool to collect and triage information, and the total amount of \textit{time} it took for participants to finish each task. For the \systemname condition, we calculate the overhead cost as the portion of the time participants spent on directly interacting with \systemname (performing wiggling gestures, interacting with the popup dialog if necessary, filtering the information clips, organizing them in the workspace view, etc.) out of the total time they used for a task (vs. reading and comprehending the web pages) \cite{liu_unakite:_2019,liu_crystalline_2022}. Similarly, in the baseline condition, the overhead cost accounts for situations where participants use the highlighting or screenshot feature to collect information, organize them in the workspace view, etc. 

\begin{figure}[t]
\centering
	\includegraphics[width=1\linewidth]{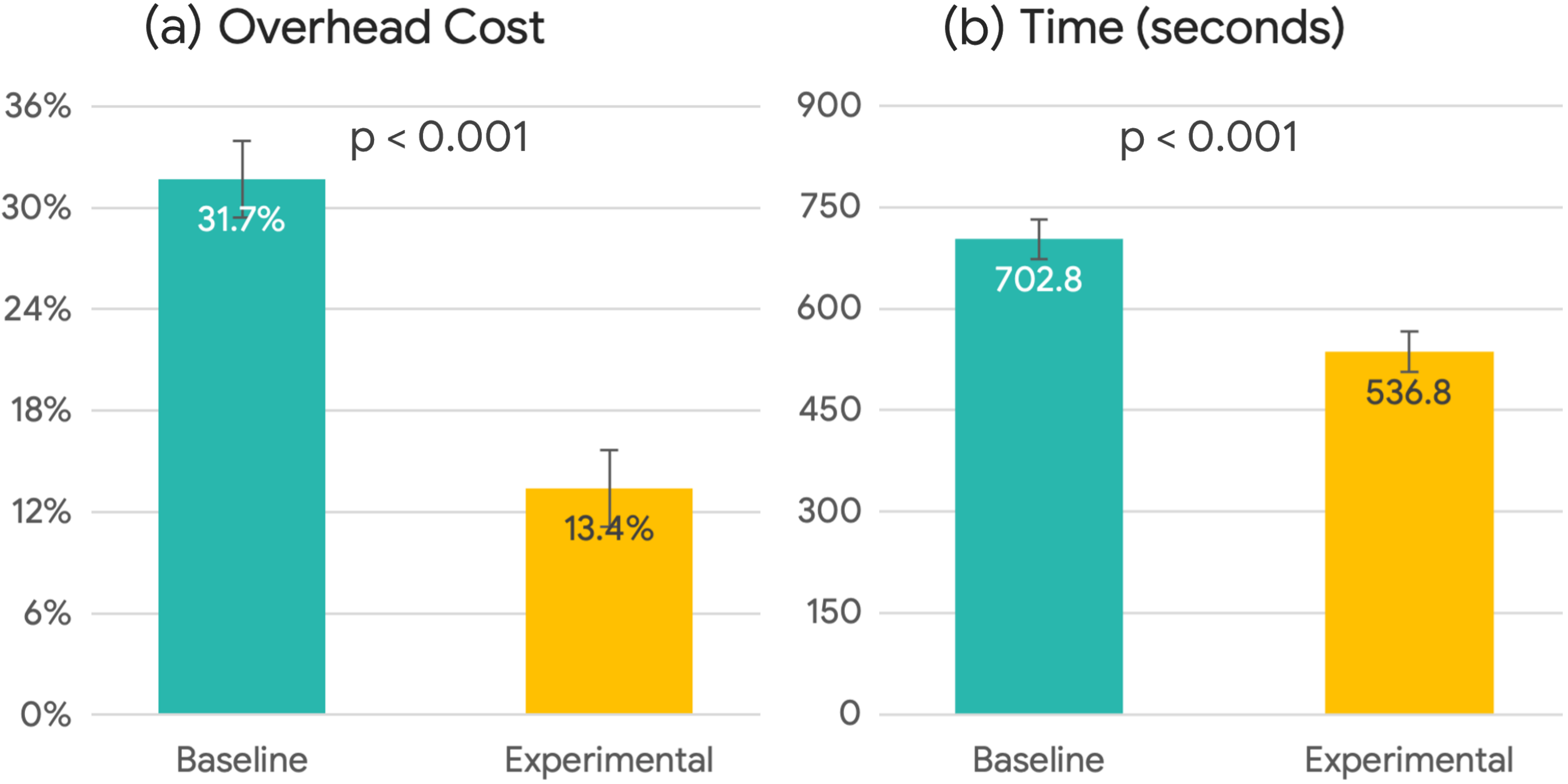}
	\vspace{-3mm}
	\caption{Using \systemname incurred significantly less overhead cost (a) and helped participants finished the tasks significantly faster (b) when compared to the baseline condition in the user study.}
	\label{fig:study-overhead-cost-and-time}
	\vspace{-3mm}
\end{figure}

We conducted a two-way repeated measures ANOVA to examine the within-subject effects of condition (\systemname vs. baseline) and task (A vs. B) on overhead cost. There was a statistically significant effect of condition ($F(1,20) = 40.7$, $p < 0.001$) such that the overhead cost was significantly lower (58\% lower, as shown in Table \ref{tab:stats-table} and Figure \ref{fig:study-overhead-cost-and-time}a) in the \systemname condition (Mean = 13.4\%, SD = 0.06) than in the baseline condition (Mean = 31.7\%, SD = 0.08). There was no significant effect of task ($F(1,20) = 0.46$, $p = 0.51$)). In addition, a two-way repeated measures ANOVA was conducted to examine the within-subject effects of condition (\systemname vs. baseline) and task (A vs. B) on task completion time. There was a statistically significant effect of condition ($F(1,20) = 20.8$, $p < 0.001$) such that participants completed tasks significantly faster (23.6\% faster, as shown in Table \ref{tab:stats-table} and Figure \ref{fig:study-overhead-cost-and-time}b) with \systemname (Mean = 536.8 seconds, SD = 67.0 seconds) than in the baseline condition (Mean = 702.8 seconds, SD = 102.9 seconds). Again, there was no significant effect of task ($F(1,20) = 0.77$, $p = 0.38$).

As the condition had a statistically significant impact on both the overhead cost as well as the task completion time (with faster completion and lower overhead cost in \systemname conditions), \systemname indeed helped participants reduce the overhead costs of collecting and triaging information and speed up their sensemaking process overall, even though the majority of their time was necessarily spent reading and understanding the material in both conditions.

Furthermore, in the post-study interview, participants overall appreciated the increased efficiency afforded by \systemname, especially using the wiggling gestures. Many (9/12) mentioned that the perceived workload to collect information that they have encountered was minimal, saying that \userquote{It felt like I didn't do anything to get those snippets into the system} (P3), and was fluid enough that it did not interrupt their flow of reading the task pages, such as \userquote{I just wiggle and move on, in fact, when I am wiggling on something, my eyes are already onto the next paragraph, no more stopping to do the regular clipping thing any more} (P11). Together with the quantitative evidence above, \systemname did offer a more fluid experience when collecting and rating information with less interruption.

\def\arraystretch{1}
\begin{table*}[t]
\centering
\resizebox{1\textwidth}{!}{%
\begin{tabular}{
>{\raggedright}p{140mm}|
p{27mm}|
p{27mm}
}
\toprule
\textbf{Question Statements} &
\textbf{\systemname condition} &
\textbf{baseline condition} 
\\\midrule

I would consider my interactions with the tool to be understandable and clear. &
6.25 (0.45) & 
6.17 (0.72)
\\\midrule

I would consider it easy for me to learn how to use this tool. &
6.42 (0.67) & 
6.33 (0.49)
\\\midrule

I enjoyed the features provided by the tool. &
6.25 (0.62) & 
6.08 (0.67)
\\\midrule

Using this tool would help make my information collection and triaging processes more efficient and effective. &
6.17 (0.39)$^*$ & 
5.75 (0.62)$^*$
\\\midrule

If possible, I would recommend the tool to my friends and colleagues. &
6.33 (0.49)$^*$ & 
5.83 (0.39)$^*$
\\\bottomrule
\end{tabular}%
}
\vspace{1mm}
\caption{Statistics of scores in the post-tasks survey. Participants were asked to rate their agreement with statements related to their experience interacting with \systemname and the baseline on a 7-point Likert scale from ``Strongly Disagree'' (a score of 1) to ``Strongly Agree'' (a score of 7). Statistics in column 2 and 3 are presented in the form of mean (standard deviation). Statistically significant differences ($p < 0.05$) through paired t-tests are marked with an $^*$. The survey questions and scales were adapted from a validated SUS scale \cite{lewis_system_2018}.}
\label{tab:survey-scores}
\vspace{-8mm}
\end{table*}

\subsection{RQ3 [Expressiveness]}

Third, we are also interested to know to what extent \systemname induces changes in people's behavior, especially given the natural extension that wiggling affords to encode priorities and valence. 

As shown in Table \ref{tab:stats-table}, participants collected significantly more information using wiggling (on average 37.8 clips, SD = 6.85) than when using the conventional selecting or screenshot workflow (on average 20.3 clips, SD = 6.68) (p < 0.01), despite spending less time on the tasks. Among the collected information clips using wiggling, 75.3\% of them were encoded with either a positive or negative valence. Similarly, participants created significantly more topics using \systemname (on average 7.83 topics, SD = 2.07) than in the control condition (on average 4.42 topics, SD = 0.90) (p < 0.01), where topics were required to be created separately in the workspace view. It is also worth noting that using wiggling to create topics (7.33 times, SD = 1.07) almost eliminated the need to separately (0.50 times, SD = 1.00) create topics (granted that most participants did at least edit the title of the topics in the popup dialog or in the workspace view to make them more succinct and easier to read).

This evidence suggests that participants indeed were able to use \systemname to externalize the perceived utility of a particular piece of information as well as their mental judgements of how it aligned with their goals in situ.


%

Furthermore, in the post-study interviews, some (4/12) participants reflected that \systemname would enable them to express their perceived utility in a way that is also useful for subsequent sorting and ranking. For example, P5 mentioned that \userquote{I really enjoyed the threading [creating topics with priorities] feature, being able to say something is important or extra important on the spot would help me stay on top of my todo list.} However, perhaps due to the limited scale of the lab study, we did not observe significant differences in the types of information participants used as topics--most of them are about the different options as well as some criteria to evaluate a product. Future and potentially larger-scale investigations are required to understand the types of information users collect using a lightweight gesture like wiggling versus using conventional capturing methods.\looseness=-1


\subsection{RQ4 [Integration]}

Last but not least, we would like to understand if the wiggle gesture would interfere with participants' normal behaviors during web browsing, such as unconsciously using the mouse pointer to guide their attention \cite{huang_improving_2012}, clicking \cite{hijikata_implicit_2004}, or scrolling (false positives). To measure this, we looked for cases where participants hit the undo button to dismiss a wiggle activation due to \systemname had wrongfully recognized some regular mouse movements as a wiggle, which turned out to be 0 across the board. This provides evidence that the wiggling gestures added by \systemname do not interfere with the existing interactions and user behaviors.

\subsection{Other Subjective Feedback}\label{sec:subjective-feedback}
In the survey, participants reported (in 7-point Likert scales) that they thought the interactions with \systemname were understandable and clear (Mean = 6.25, SD = 0.45), \systemname was easy to learn (Mean = 6.42, SD = 0.67), and they enjoyed \systemname's features (Mean = 6.25, SD = 0.62). In addition, compared to the baseline condition (Mean = 5.75, SD = 0.62), they thought using \systemname (Mean = 6.17, SD = 0.39) would help make their information collection and triaging processes more efficient and effectively (p = 0.017), and would recommend \systemname (Mean = 6.33, SD = 0.49) over the baseline version of \systemname (Mean = 5.92, SD = 0.29) to friends and colleagues (p = 0.007), both differences were statistically significant under paired t-tests. Details of the survey questions and scores are presented in Table \ref{tab:survey-scores}\looseness=-1.

In addition, some participants reflected on the playfulness and attractiveness of the wiggle interactions and how it encouraged them to collect information compared to what they normally have to go through. For example, P8 said: \userquote{It's fun, you know? I didn't quite believe it at the beginning, but it actually made grabbing stuff so much fun}, and P1 suggested that \userquote{somehow with this, I don't think going through something that I'm not familiar with would be as daunting as it used to be}. Four of the participants even went on to ask when \systemname will be released publicly so that they could use it for their own work and personal tasks, and wondered if they could customize the system, such as by \userquote{writing some sort of plugin, like the one I wrote for Obsidian \cite{obsidian_obsidian_2022}, to map the different directional swipes to what I want depending on the situations that I'm in} (P11).

\section{Limitations}\label{sec:limitations}
One potential limitation to wiggling is the suitability of its rapid back-and-forth movements to user populations with motor impairments or advanced age, for example, users with hand tremors. There are several ways in which wiggling might be more suitable than expected or relatively easily adapted to such populations. First, since wiggling uses the initial mouse location as its selection anchor, a user can take their time adjusting to arrive at the correct area (which would still require less accuracy than traditional highlighting). Once there, they could initiate selection without clicking, which could address mouse slip while clicking, a common problem with advanced age or motor impairment \cite{trewin_developing_2006,fan_eyelid_2020,fan_eyelid_2021}. If issues with tremor lead to lower accuracy, one approach that might be investigated is smoothing mouse movement using generative models trained on a user's individual behavior (e.g., \cite{williams_real-time_2016}). More generally, additional research is needed to understand the suitability of wiggling across a variety of user capabilities, contexts, and devices \cite{li_freedom_2022,li_i_2021}.

Our lab study had several limitations. Given the short amount of training and practice time, some participants might not have been able to fully familiarize themselves with the wiggle-based collection and triaging techniques offered by \systemname. The study tasks and topics might not be the ones that participants typically encounter, and therefore they may not have sufficient motivation or background context as in real life. However, we attempted to mitigate these risks through carefully preparing the study setup: (1) we chose the training and real study tasks based on actual product comparison topics that people are faced with; (2) we had participants practice using \systemname as well as its baseline version for each condition simulating what they needed to do for the real tasks, and (3) we provided participants with ample amount of background information to help them get prepared. 
We would like to further address these limitations in the future by having participants use \systemname for their own work and personal tasks and projects, which would presumably fuel them with the necessary motivation and context and engage with \systemname in a more organic way.


While sensemaking in various domains might exhibit different characteristics and therefore lead to different information foraging behavior patterns, we chose both the study tasks to be in the domain of comparison shopping to at least make sure that the tasks are roughly of equal difficulty. In addition, product comparison shopping embodies many of the common sensemaking properties and needs that people have, for example, it is information dense so that users would potentially have to read and process lots of information and collect quite a few items, and users would often have to interpret the information based on their own goals and context, so that there is a need for them to externalize their mental context alongside the collected information. Nevertheless, we would like to address this limitation by evaluating \systemname in a variety of domains where sensemaking usually occurs, such as students conducting literature reviews, patients researching medical diagnoses, and programmers learning unfamiliar APIs.

There was also a risk of participants already being familiar with a topic, such as an expert photographer doing task (A). However, in the post-study interviews, we confirmed that none reported having extensive experience or expertise in any of the task topics.

Finally, due to the limited set of capabilities of the \systemname mobile application and the similarity of features with its desktop counterpart, we did not evaluate the mobile app in our lab study, and therefore could not directly compare the wiggle-based collection and triaging techniques for sensemaking to a baseline. Informally in our pilot testing, using wiggling to collect information and optionally encode a positive or negative valence was much faster and more convenient than any common information capturing methods that people currently use on mobile devices, such as copying and pasting text and taking screenshots or photos \cite{swearngin_scraps_2021}. Nevertheless, we would like to evaluate the wiggle-based techniques and \systemname for mobile in a formal lab study in the future.

\section{Future Work}


An essential goal of this work is to explore ways to enable people to focus on reading and comprehending actual content rather than splitting attention on the mechanics of collecting information as well as externalizing their mental context. However, prior research \cite{council_how_2000,bates_design_1989,kerne_using_2014} has suggested that there is a higher likelihood for people to recall and trust the information if they consciously spend time collecting and synthesizing it with the existing information. This raises an interesting tension and trade-off between pursuing low-cost interactions for information capturing and triaging versus consciously collecting and synthesizing the encountered information --- future research would be required to examine the long-term effect of using lightweight systems like \systemname on people's learning outcomes as well as decision results in various kinds of sensemaking scenarios.

Research on activity-based computing \cite{dearman_its_2008,brudy_cross-device_2019} has suggested the benefits of granting users access to their information repository as well as the ability to perform tasks across multiple devices. While the current implementation of \systemname mobile application does enable users to collect information and triage it with valence as well as to review their collected information, extending it to support more complex operations such as creating and curating topics could be an interesting direction for future work. We anticipate two challenges: (1) how to reasonably leverage the limited screen real estate to design user interfaces and interactions that feel native to a mobile device while also being functionally lightweight and intuitive enough to perform, and (2) exploring the realistic role of mobile devices in the sensemaking ecosystem \cite{williams_mercury_2019,iqbal_multitasking_2018}, i.e., striking a balance between pursuing feature parity with the \systemname desktop counterpart and designing to specifically support practical use cases (e.g., reviewing and triaging information during a commute).


Though we extended the wiggle gesture to support encoding the valence and priority of information, we envision a larger design space where different aspects and properties of the wiggling movement could be mapped to various functions to increase its expressiveness and utility. For example, the \textit{speed} of each movement, the \textit{duration} of the total movement, or the \textit{size} of the gesture (currently mapped to target size selection) are all continuous measures that may intuitively map to various qualities that could be used in interactions, such as uncertainty or confidence towards some content. Building on what we mentioned in section \ref{sec:subjective-feedback} and \ref{sec:limitations}, future work could investigate: (1) ways to support customizing the mapping of the different aspects of a wiggle according to users' preferences and sensemaking scenarios, and (2) intelligently learning and adapting to users' needs and habits over time, such as re-calibrating the recognition software to account for individuals' cognitive and physical conditions \cite{trewin_developing_2006}.

Last but not least, we envision a future where the wiggling technique as a new class of interaction can be extended and applied to tasks and applications other than sensemaking. On the one hand, a consistent application scenario envisioned by participants involved using wiggling for repeated selection, extraction, and annotation of various types of data as part of a data processing task, such as aggregating recipes online before going shopping, or saving specific torque numbers of fasteners while working on a car.
On the other hand, wiggling might also be used for many other system-level behaviors as it does not conflict with most traditional selections or gestures. For example, wiggling with popup or cross through menus could be applied for quickly modifying device settings on the fly, such as screen zoom or brightness. Additionally, wiggling could serve as an initial activation for a much more expressive set of gestures, or even summon an intelligent agent (such as a voice assistant) that can perform a complex action on the specified item.

\section{Conclusion}
This work explored a new interaction technique called ``wiggling,'' which can be used to fluidly collect, organize, and rate information during early sensemaking stages with a single gesture. Nowadays, people face the challenge of capturing the information they find for later use as well as externalizing their mental judgement about its valence and priority during many online learning and exploration tasks, such as conducting comparison shopping, researching medical diagnoses, and searching for code snippets. While current select-copy-paste-based approaches incur a high cost when doing so, the wiggling technique afforded by the \systemname system discussed in this paper only involves light-weight back-and-forth movements of a pointer or up-and-down scrolling on a smartphone, which can indicate the information to be collected and its valence in a single gesture, and does not interfere with other available system interactions. In addition, we envision an extensive design space where the wiggling interaction can be adapted to support a wide range of custom actions beyond sensemaking in the future.

\begin{acks}
This research was supported in part by NSF grants CCF-1814826 and FW-HTF-RL-1928631, Google, Bosch, the Office of Naval Research, and the CMU Center for Knowledge Acceleration. Any opinions, findings, conclusions, or recommendations expressed in this material are those of the authors and do not necessarily reflect the views of the sponsors. We would like to thank our study participants for their kind participation and our anonymous reviewers for their insightful feedback. We sincerely thank Jinlei Chen, Tianying Chen, Haojian Jin, Toby Jia-Jun Li, Franklin Mingzhe Li, Haitian Sun, Jiachen Wang, Eric Yiyi Wang, Ziyan Wang, Bella Kexin Yang, and Zheng Yao for their constant support, especially during the COVID-19 pandemic.
\end{acks}

\bibliographystyle{ACM-Reference-Format}
\bibliography{aaareferences}

\newpage
\appendix

\section*{APPENDIX}
\setcounter{section}{1}

\subsection*{\basesystemname Content Clipping}

For clipping, \basesystemname offers two methods:

\begin{itemize}[leftmargin=1.2em]
    \item \textbf{Clipping text}: First, use the cursor to select the desired content (Figure \ref{fig:skeema-basic-clipping}a in the appendix) in the conventional way, then click the clipping button (containing the \basesystemname chrome extension icon, see Figure \ref{fig:skeema-basic-clipping}a1) that popups to collect the selected texts into an information card in the holding tank. After that, there will be a popup dialog (Figure \ref{fig:skeema-basic-clipping}c) on the upper right corner of the screen to indicate success as well as allow users to externalize their mental context such as valence judgement into the notes field associated with the collected content.
    \item \textbf{Clipping screenshot}: To collect images and other types of non-text content, users can use the screenshot feature: they need to click the screenshot button (Figure \ref{fig:skeema-basic-clipping}b1) on right edge of the browser window to enter the screenshot mode, press and hold down the mouse left button to drag out a bounding box around the desired content, and release the left button. Then the screenshot will be saved into the holding tank as an image, followed by the popup dialog (Figure \ref{fig:skeema-basic-clipping}c).
\end{itemize}

\begin{figure}[h]
\centering
	\includegraphics[width=0.9\linewidth]{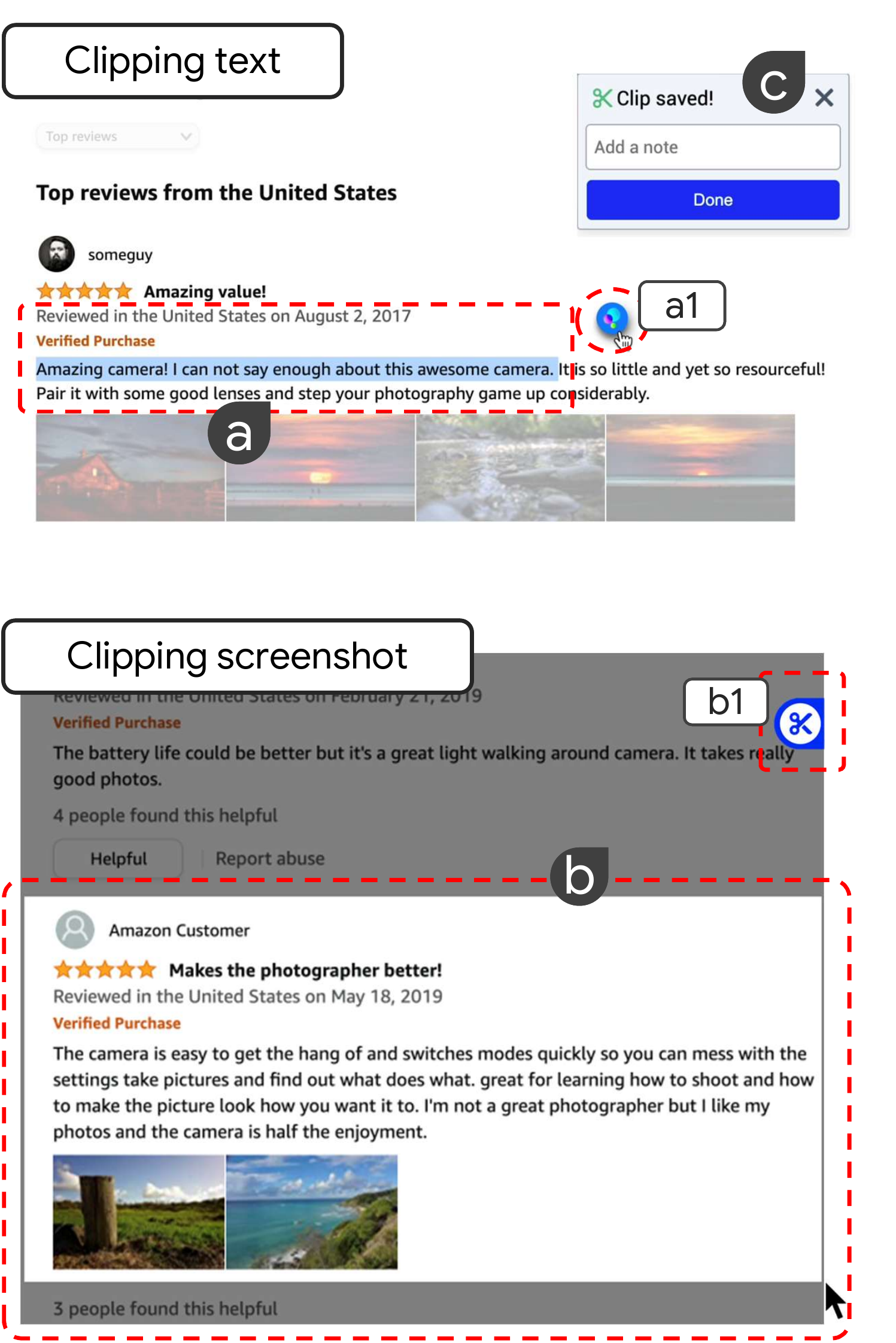}
	\vspace{-3mm}
	\caption{Two types of information clipping mechanisms that \basesystemname supports: clipping text (top) and clipping screenshot (bottom).}
	\label{fig:skeema-basic-clipping}
\end{figure}

\end{document}